\documentstyle{article}

\def\be{\begin{equation}}
\def\ee{\end{equation}}
\def\bea{\begin{eqnarray}}
\def\eea{\end{eqnarray}}

\newtheorem{theorem}{Theorem}
\newtheorem{lemma}{Lemma}
\newtheorem{remark}{Remark}

\title{{\bf Integrable deformations of systems on graphs with loops}
\vspace{.5cm}}
\author{{\bf L. O.
Chekhov}\thanks{E-mail: \ chekhov@mi.ras.ru. The work is supported by
the RFBR Grant No.~01--01--00549 and by INTAS Grant No.~99--01782.} 
\date{ } \\
{\it Steklov Mathematical Institute, Moscow, Russia.}}

\begin{document}

\maketitle

Studying spectral properties and scattering processes
for the Schr\"odinger-type
operators on infinite trees~\cite{Nov1}~\cite{Nov} has provided 
many unexpected results including the relation between the scattering
($S-$) matrix on an infinite tree with cycles and Artin--Selberg
$L$-function~\cite{Ch}. However, no reasonable integrable
deformation of second-order operators that resembles that for
the lattices~$\bf Z$ ($(L,A)$-pairs of the Toda chain type)
and~${\bf Z}_2$ ($(L,A,B)$-triples)
(see~\cite{NovDyn}) had been constructed.

In~\cite{KN}, Krichever and Novikov had constructed the deformation
of arbitrary fourth-order symmetrical operator $L$ in the form of the
$(L,A,B)$-triple, $\dot L=LA-BL$, for the
$T_3$ graph (infinite homogeneous tree with vertices of valence three).
This deformation preserves the spectral
data of the zero level $L\psi=0$.

In~\cite{ChP}, the Krichever--Novikov integrable system was generalized
to the case of an arbitrary (inhomogeneous) tree graph~$T$.
An arbitrary graph~$\Gamma$ with finite number of cycles can be
innambiguously (up to isomorphisms) presented as the quotient of the universal
covering tree graph~$T$ by the action of a finitely generated
subgroup~$\Delta$ of the total group of symmetries of the tree graph~$T$,
$\Gamma=T/\Delta$.

We consider operators $L$ acting on the space of functions
$\psi_P$ on vertices of a graph~$\Gamma$. The distance
$d(X_1,X_2)\in {\bf Z}_+\cap\{0\}$ on a tree graph
is measured in number of edges entering the path connecting the
vertices~$X_1$ and~$X_2$. The order of equation
$L\psi=0$ where $(L\psi)_P=\sum_{X}^{}b_{P,X}\psi_X$ is the maximal diameter
$\max_P d(X_1,X_2):\, b_{P,X_1}\ne 0, b_{P<X_2}\ne 0$ or $b_{X_1,X_2}\ne 0$.
In the general graph, paths connecting two vertices $X_1$ and $X_2$ are not
unique. We can then consider the theory on the covering tree~$T$ claiming
all functions $\psi_X$ and coefficients of the operators~$L$, $A$, and~$B$
to be periodic with respect to the action of the symmetry subgroup~$\Delta$.
On the factorized graph~$\Gamma$, this means that we associate coefficients
of the operators with {\it paths} of the corresponding lengths connecting
preimages $\tilde X_1,\tilde X_2\in T$
of the corresponding points $X_1,X_2\in \Gamma$.

We first consider an arbitrary graph without loops.

\begin{theorem}\label{th1}~\cite{ChP}
Real self-adjoint operator $L$ of the fourth order on a graph
$\Gamma$ without loops admits an isospectral deformation of a zero energy
level $L\psi=0$ having the form of the
$(L,A,B)$-triple $[B=A^t]$,
\be
\dot L=LA+A^tL,
\label{a.1}
\ee
where $A$ is the second-order operator,
\be
(A\psi)_X=
\sum_{X':\,d(X,X')=1}^{}a_{X,X'}\psi_{X'}+s_X\psi_X,
\label{a.2}
\ee
iff the operator $L$,
\be
(L\psi)_X=\sum_{X'':\,d(X,X'')=2}^{}b_{X,X''}\psi_{X''}+
\sum_{X':\,d(X,X')=1}^{}r_{X,X'}\psi_{X'}+\rho_X\psi_X
\label{a.3}
\ee
(such that all $b_{X,X''}>0$) can be presented in the form
\be
L=Q^tQ+u_X,
\label{a.4}
\ee
where $Q$ is a second-order operator
\be
(Q\psi)_X=
\sum_{X':\,d(X,X')=1}^{}q_{X,X'}\psi_{X'}+v_X\psi_X.
\label{a.5}
\ee
\end{theorem}

In the case where $b_{X,X''}\equiv b_{X'',X}>0$,
the coefficients $q_{X,X'}$ are inambiguously determined
by the coefficients $b_{X,X''}$ and vice versa. In particular, for
arbitrary three points $X_1,X_2,X_3$ such that $X_1\ne X_3$ and
$d(X_i,X_{i+1})=1$, \ i=1,2, $b_{X_1,X_3}=q_{X_1,X_2}q_{X_3,X_2}$.
The coefficients $v_X$ depend on a free parameter---the value of
$v_{X_0}$ at the given point $X_0$ (the center).
[An arbitrary fourth-order operator $L$ (\ref{a.3}) on the tree
$\Gamma_3$ admits representation (\ref{a.4}) and, hence, isospectral
deformation (\ref{a.1})~\cite{KN}.]

The coefficients
$a_{X,X'}$ of deformation operator (\ref{a.2}) are inambiguously expressed
(up to a total multiplicative constant) via
$q_{X,Y}$. For arbitrary three vertices
\be
X_1,X_2,X_3: \ d(X_i,X_{i+1})=1, \ i=1,2,
\label{a.6}
\ee
we obtain\footnote{It was a misprint in this formula in~\cite{ChP}.}
\be
\frac{a_{X_2,X_1}}{q_{X_1,X_2}}=\frac{a_{X_2,X_3}}{q_{X_3,X_2}}\ \hbox{and}\
a_{X_3,X_2}=-a_{X_2,X_1}\frac{q^2_{X_2,X_3}}{q_{X_1,X_2}q_{X_3,X_2}}
\label{a.7}
\ee
[in particular, when $X_1=X_3$ we obtain
$a_{X_1,X_2}=-a_{X_2,X_1}q^2_{X_2,X_1}/q^2_{X_1,X_2}$].


Let $\{X'_j\}$ and $\{X''_i\}$ be the sets of vertices that are neighbor
to the respective vertices $X_1$ and $X_2$ but do not coincide with $X_1$
or $X_2$. Explicit formulas for the deformation read
\bea
\dot b_{X_1,X_3}&=&r_{X_1,X_2}a_{X_2,X_3}+a_{X_2,X_1}r_{X_2,X_3}+
b_{X_1,X_3}(s_{X_1}+s_{X_3})\nonumber\\
\dot r_{X_1,X_2}&=&\sum_{X''}b_{X_1,X''}a_{X'',X_2}+
\sum_{X'}b_{X_2,X'}a_{X',X_1}+\rho_{X_1}a_{X_1,X_2}+\rho_{X_2}a_{X_2,X_1}
+r_{X_1,X_2}(s_{X_1}+s_{X_2}),
\label{a.8}\\
\dot\rho_{X_1}&=&\sum_{\{X'\}\cup X_2}r_{X_1,X'}(a_{X',X_1}+a_{X_1,X'})
+2\rho_{X_1} s_{X_1}.
\nonumber
\eea

\begin{lemma}~\cite{ChP}.
For arbitrary tree graph,
deformation (\ref{a.8}) by virtue of
(\ref{a.7}) preserves conditions
(\ref{a.6}) (and, hence, factoring condition (\ref{a.4}).
\end{lemma}

We now consider arbitrary graph with vertices of valence three and finite
number of loops.

\begin{theorem}\label{th2}
For arbitrary graph $\Gamma_3$ with vertices of valence three and arbitrary
number of loops, deformation {\rm(\ref{a.8})} preserves the periodicity
properties iff for any closed cycle with the consequtive (cyclically ordered)
vertices $\{X_i\}_{i=1}^n$, \ $X_{n+1}\equiv X_1$, the two conditions
\bea
\hbox{{\rm(a)}}
&{}& \prod_{k=1}^{n}(-)\frac{q_{X_k,X_{k+1}}}{q_{X_{k+1},X_k}}=1
\quad \hbox{\rm (cocycle\ condition,\ see\ \cite{KN})},
\label{a.9}\\
\hbox{{\rm(b)}}
&{}& \sum_{i=1}^{n}r_{X_i,X_{i+1}}
\prod_{k=1}^{i-1}(-)\frac{q_{X_k,X_{k+1}}}{q_{X_{k+2},X_{k+1}}}=0
\label{a.10}
\eea
are satisfied.
\end{theorem}


\begin{remark}
{\rm
The proof is the lengthy but direct calculation using formulas
(\ref{a.7}), (\ref{a.8}). Note that the time derivative of condition (a)
is twice the condition (b), and the only nontrivial part is to prove that
the time derivative of condition (b) necessarily
vanishes providing both conditions (a)
and (b) are satisfied. Therefore, no new conditions arise. Because
both condition (a) and condition (b) 
satisfies the cocycle property, it suffices
to impose these conditions only on base cycles of a graph.
}
\end{remark}

\end{document}